**Slow electronic rearrangement kinetic involving O$^-$ after shot quenching from >600K of YBa2Cu3Oy**

H. Oesterreicher, Department of Chemistry, UCSD, La Jolla, CA  92093-0506

Analysis of kinetic literature suggests several electronic rearrangements above 600K involving O$^-$ (subperoxides) in YBa2Cu3Oy. Slow cooling to 260K produces a Tc=50K level with conventional plane and cell volume contracted varieties (P-V-), based on plane oxidizing (4) where (n) denotes local O coordination. When these materials are quenched from <600K, the subsequent changes at 298K of structural parameters or Tc have activation energies E=92kJ and A=1.0x10$^{-12}$s. The changes saturate at temperatures where they are faster than the time constant t of quenching. This temperature is extended to 473K with SQ, indicating t=0.1s. SQ of y near 6.44 from >600K leads to increased axial ratios, plane and V expansion (P+V+), in a process which is complete near 670 and extending to 950K. P+V+ effects can far exceed P0V0 of the respective semiconductor. The relative stability at elevated temperatures indicates another intrinsically slower transformation mechanism of primarily electronic nature with E=280kJ, involving O$^-$. V+ and its Tc=100K level, is ascribed to (3) with doped charge on apical O$^-$, acting as a plane expander and n-doper. Similar slowing of kinetic through O$^-$ also hold for the more complex range of Tq>950K. It can produce a c-axis contracted non-superconductor (P+C-) and indications for elevated temperature superconductivity (ETS) with Tc>150K*. Kinetic data are embedded into general thermodynamic arguments concerning peroxide stability belts.

**1. Introduction:**
     We analyze existing literature concerning the potential for kinetically stabilizing unusual doping types in YBa2Cu3Oy=6+z near O half filling (123/6.5) [1]. The existence of complex high temperature modifications on shot quenching (SQ) indicates inherently slow transformation kinetic (high activation energy E) to low temperature modifications. This is here taken as a telltale sign for phenomena involving O$^-$ and the need for an electronic rearrangement during transformation. Shining light on the electronic nature of some 123/6.5 has gained new relevance due to the discovery of signs for elevated temperature superconductivity (ETS), induced as a fleeting state through short infrared laser pulses [2,3]. These studies showed indications for Tc=552K* where stared values denote ETS, which has been argued to be connected with additional O$^-$ pairs on chains and attendant 3D effects [4] with the pairs connected with the empty chain site [5]. In addition, new electronic types include plane and cell volume expanded types (P+V+) due to presumed n-doping on planes [6-11] as well as indications for ETS were also found on SQ from >600K. They can show Tc levels of 100 and 200K* (as minority component) [6,7]. For understanding these phenomena it is important to investigate transitions in the context of structural and electronic reordering. Accordingly arguments will be presented that the often-observed O$^-$ through spectroscopy [12] can be generated in the self-doping process when materials are quenched through elevated temperature belts. O$^-$ can serve as the common denominator to uncommon phenomena both electronic and kinetic.
     This is contrasted to slow cooling to 260K, which produces conventional plane and cell volume contracted varieties (P-V-), based on plane oxidizing (4) where (n) denotes local O coordination, in this case on the chain Cu site. The corresponding structure is denoted O42 (also denoted orthoI or OI). The crossover in this metal ligand charge transfer in the transition from V+ to V- has been shown to happen around 600K with further complexity at yet higher temperatures [1]. Around 673K the quenched material optimizes V+P+ properties and consists primarily of (3) on the chains with small amounts of (2) denoted as O32 or O3 for simplicity. SQ from >1050K can produce several semiconductors at cell volume V0, between V- and V+.

It was realized in the pioneering studies that O42, observed on oxidizing slow cooling, was a low temperature modification. O42 could change slightly when quenched from moderately higher temperatures to modifications involving increasing amounts of (3) at constant y (O243). As a result transitions from lower Tc to higher values were routinely observed on room temperature annealing of moderately fast quenched materials due to the transition from O243 back to O24. [1, 6-11, 13-17]. In subsequent studies it was found that this represented only part of a larger phenomenology of transformations at elevated temperatures [1, 6-11]. This initially missed, and most intriguing aspect of phenomenology sets in at >600K. It can be observed on extending the speed of quenching techniques and by engineering increases in activation energy of the transition reaction through chemical means (partial substitutions). Shot quenching with small sample size of 0.2g shortened quenching time to 0.1s. Indications for strong V+P+ effects appeared for several rare earth and Y analogs near experimental y=6.46, which was assumed to reflect an ideal plane bond order [18-20] at y=6.444 with doped charge concentration on planes of c=0.222. The V+ effects started to pick up at quenching temperatures, corresponding to the subperoxide stability belt [21,22], and maximized at 670K with volume increases of up to 2.7% over V- [6]. In some instances they were coupled with a doubling and quadrupling of Tc. Minority components with Tc=200K* level are here related to ETS. Around 1050K a phase with strong plane expansion and c-axis contraction but volume corresponding to the semiconductor (V0) was found and designated C- phase of C-V0 type [23,24].

These studies were stimulated by ones of the effects of partial transition metal substitutions for Cu. The resulting micro-cluster formation [25-32], leading to some reductions in Tc and rise in magnetic hardness, was also found to reduce the speed of transformation reactions from the metastable high temperature O3 modifications to the conventional O42 orthorhombics. On partial Al substitution for Cu, chain length was also reduced [33]. By utilizing this slowing of transformations, the monitoring of fast transitions was facilitated. The main part of this work involved YBa2Cu3-xNixOy with x<0.05. While strong cell volume expansions were found also with x=0 and with several rare earth analogs, the increases in Tc when calibrated to YBa2Cu3-xNixOy with x=0.0 were so far only found with the Y analog. In YBa2Cu3-xNixOy considerable complexity is encountered concerning the response to oxygenating (low temperature equilibration) or reducing preparations (shot quenching) termed Ko phenomenology. Besides increases in Tc, also absence of Tc at high y is observed and these materials show large amounts of (3) and resistance to re-oxygenation. In fact, Ko phenomenology can be attributed to the competition of local environments of (3) and (4) with their, under certain circumstances, inherently different doping mechanisms based on O$^-$ versus straightforward ionic metal charge storage respectively.

General stability criteria for O ordering were given in terms of O-O potentials [34,35]. General stability criteria for structures in question were presented in terms of thermodynamic redox potential [22] vis-à-vis O uptake including peroxides and metastability. n-doping in cuprates is reviewed in [36].

**2. Results**

A challenge to the materials synthesis field lies in the exploitation of complexity in the phenomenology of YBa2Cu3Oy near O half filling at elevated temperatures. Selected electropositive elements are reacting with O to form peroxides and derived principles, based on bonds involving O$^-$ in this range. Examples are BaO2 or BaCuO2.7. Not surprisingly, derivatives of this bonding principle characterize also 123/6.5. It will be argued that they are examples of subperoxides that do not liberate O2 in acid. As electronic rearrangements can be characterized by slow transformation kinetic, subperoxide based materials have a chance to either survive quenching to lower temperatures, or come into existence during quenching and survive as transitional principles.

Table1 presents a summary of temperatures for experimentally accessible phases in 123/6.5 as usually encountered: T32 (1100K), T34C- (1050K), amorphous (970K), O3 (950-670K), O342 (670-600K), O423 (600-260K). Thermodynamic stability is a more involved question, hidden behind complex transformation kinetic. Such transformations between various phases occur in 123/6.5 after various heat treatments in a way dependent on thermal sequencing. Prior thermal history therefore determines the further development of structural and electronic properties within several branches as is also presented in Table1. The observation of alternative high temperature modifications at room temperature indicates that these transform to the conventional O24 unusually slowly due to high activation energies. This indicates that they are of a different electronic nature.

Electronically speaking, one notices that the end members, T32 and O24, are semiconducting and self-doped through an ionic metal respectively. Between these end members lie regions incorporating $O^-$ in a way to be further elaborated on below. In addition, amorphization is observed at 970K, potentially indicating a range of insipient decomposition into metal oxides and peroxides such as $BaCuO_x$ with x=2.5 to 2.7, or $BaO_2$. This has to do with the higher overall O content (e.g. 8.5) compared with 123/6.5 that can lead to higher overall exothermic heat of formation. The presence of a competing peroxide stability belt points generally into the direction that $O^-$, observed in high temperature modifications, has properties reminding of peroxides. We consider these properties to encompass charge equilibration between $O^-$ and $O^{2-}$ within selected O aggregates such as the apical system. These units are referred to as subperoxides $O_q^{1-2q}$. Hallmarks of the postulated subperoxides include structural fingerprints (cell volume and plane expansion as well as altered axial ratios), absence of O2 production in acid, presence of (3) or (2) and unusual superconducting properties indicating potential for massively increased pair concentration.

In selecting a synthesis path, a main difference pertains to T23 (majority (2)) and O32 parentage (majority (3)), gating development into orthorhombics based on (4) and (3) respectively, the later being potentially related to $O^-$. T23, obtainable on annealing (a) at >1100K (y<6.25) can develop, on slow oxygenating cooling (SC), into conventional O24 of P-V- type (Tc=50K level). O24 is based on oxidative self-doped $(5)^{Ox}$ through charge equilibration with $(4)^{Red}$ (bond order arguments indicate that the magnitude of self-doping corresponds to ¼ of the available dopant so that for z=0.5 with 50% (4), c=z/4=0.125). By contrast, shot quenching (SQ) y>6.4 (673 to 950K with ideal y=6.44) can develop into phenomenology of O32, containing doped oxidized (Ox) apical $O^-$. This $^cO^{Ox}$ on (3) can lead to P+V+ effects, indicating charge equilibration leading to $(5)^{Red}$ (Tc=100K level). We assume $O^-$ is present in the form of subperoxides. As occasions arise where we don't have detailed knowledge of the in situ situation we denote parent high temperature modification by symbol p such as V+p. These parent modifications can develop their intriguing electronic and structural properties through thermal stress on SQ.

A further distinction pertains to transformations of O24 and O32 derived materials. These transformations can happen on aging at room temperature or at synthesis conditions. The transformations within O24 have lower activation energies E. When O24 is held up to 600K and shot quenched it develops O243 with non-doping (3). O243 is transforming back to O24 with E=92kJ and $A=1.0 \times 10^{-12}$s after SQ to 298K. By contrast, varieties based on $O^-$ such as O32, obtained by annealing at 950K, transform slowly to O24 with E up to 280kJ. This is extracted from the observation of a change from V+ to V- within an hour annealing at 950K. The reaction is too slow to be practically observed at 298K.

Thermodynamic parameters are determined according to $t=Ae^{E/kT}$. The transition within V- corresponds to materials initially prepared by very slow cooling. Accordingly the smallest cell volumes are obtained at 173.0$A^3$ with saturation of Tc and V- effects near 260K. Samples are then shot quenched in the temperature interval 320-473K and structure and Tc measured.

Phenomena saturate for >473K as at this temperature the shot quench has t=0.1s, limiting observation of this type of transformation at higher temperatures. Values of E=92kJ are typical, as is A=1.0x10$^{-12}$s. For 298K, t=300min is typical with these values. One also notes a similarity of E with values of O diffusion, indicating that these processes are diffusion controlled. Values of E increase to 96kJ on partial Ni substitution according to YBa2Cu3-xNixOy with x=0.020.

While V increases mildly up to Tq=600K, due to the incorporation of non-doping (3), it does so at a much faster rate for Tq>600K maximizing near 670K. At that point V can reach values of 177A$^3$. This V+ is roughly symmetrically disposed from V0, the value corresponding to the conventional semiconductor, compared to the negative deviation in V- from V0. Also values of c/p, with p=3(a+b)/2, increase in a characteristic way. We assume that these changes have to do with incorporation of O$^-$. Changes from this state to O24 are of an electronic nature and therefore much slower (higher E). One also notes that E has to be increased in order to survive SQ with t=0.1s from Tq>470K. Taking lnt=lnA=e/kT one can write, with A=1.0x10$^{-12}$s, T=E/210 for SQ with t=0.1s.

The above summarizes the difference of transformations within V- and V+ respectively. They encompass SQ from 600K to 260K for inter V- transformations as the conventional part, and the range of SQ from around 670K for V+ to V- transformations as a new variation of the theme. However, mixed transitions are also possible. Examples obtain for SQ in the range from 670 to 600K. In this case the two independent transformations are superimposed. At 298K one notices practically only the transformation O243 into O24 as it has low E. However, the part involving O$^-$ is maintained and manifest as relatively large V and c/p.

A dramatic electronic change happens accordingly for Tq>600K. Above this temperature one can assume that the equilibrium $M^{n+}L^{2-} = M^{(n-1)+}L^-$ starts to become shifted to the right. These are the materials that show O$^-$ in spectroscopy even at 298K. One should therefore keep in mind that at the quench temperature for achieving V+ effects, materials need not, and probably do not, display V+ or related effects in situ. Rather, we assume that they are generated anew by the tension/compression upon quenching from the known types of argyle-O3 or herringbone-T3 semiconducting types. Generally, the windows where unusual effects are observed remind of classical peroxide stability belts.

The situation becomes more complex for Ta>900K as the V+ to V- transformation moves into the observable temperature and time window. Examples are: When a conventionally prepared material is quenched into liquid N2 from 1080K and subsequently air annealed at 950K for 10 min, followed by quenching, a relatively V expanded orthorhombic is observed. However, after 2 h at 950 K a cell volume contracted orthorhombic obtains (the direct transformation from V+ to V- at 950K with t=1h has then a high E=280kJ). The Tq=1080K material held for 10 min at 970K does not produce resolved X-ray diffraction peaks. Initial calcining or re-calcining of HNO3 dissolved conventional material at 1050K leads to the C- phase of T34 type. The preparation at 970K is therefore leading to some sort of amorphization or micro-decomposition, either due to a transition between relatively unrelated C- phase of T3 parentage and O3 respectively or to the inherent metastability at that temperature vis-à-vis decomposition into metal oxides and peroxide BaCuOx that represent the thermodynamically stable state at higher O uptake. In situ studies indicate the boundary between T and O at 940K.

The C- phase as obtained by slow cooling after calcining at the gating temperature of 1050K has apparently yet higher E according to E=296T for t=3600s. Its relation to ETS has been outlined earlier and above. However, its O$^-$ involvement of appears to be of a more massive type potentially involving local peroxide bonds rather than subperoxidic ones. This would suggest that its stability range is closer to V+ O32, or Ta<1050K.

The critical composition for V+ O32 at y= 6.44 has been connected with bond-order stability of a pair within a kernel of 3ax3b, resulting in an electron concentration of c=2/3x3=0.22. The electronic nature of the delocalized charge may have aspects in common with solvated electrons or electrides [37,38]. In the self-doped case the plane is the location for a correlated crystal of moving electrons.

## 3. Discussion

This study extracts new information from existing literature about inherent stability in a thermodynamic and kinetic sense for YBa2Cu3Oy near oxygen half filling. Generally, materials are thermodynamically stable near synthesis conditions of the semiconductor. On cooling, with the chance of further O uptake, they become metastable vis-à-vis peroxides such as BaCuO2.7 and metal oxides. This has to do with the increased oxygen uptake of decomposition products over the superconductor. However, this decomposition reaction has high activation energy and therefore a series of metastable states can survive cooling, although amorphous states observed on holding materials at 970K may indicate the beginning of this decomposition. On careful slow cooling procedure through this experimental window of instability one will obtain O24 as the eventual metastable survivor at room temperature. Another exception to decomposition, at higher temperature, is the C- phase obtained after initial calcining at 1053K. It also appears to contain peroxidic $O^-$ and relatively high y.

The range around 940 to 950K is characterized by the transition from tetragonal to orthorhombic on in situ studies. When these O32 materials near ideal y=6.44 are SQ one can find strong cell volume expansions at 298K. A further equilibration near 670K can further accentuate these V+ effects that can also lead to a doubling of Tc to a 100K level. These materials can also contain minorities with a Tc=200K* level hinting at an ETS state. We have here taken a particular experiment to indicate that the transformation of O32 to O24 has a strongly increased E=280kJ by comparison with the transformation within O24 (O243/O24 with E=92kJ).

The relative stability of these SQ materials is of relevance to the possible stabilization of an ETS state. A relation of this state to V+ O32 is assumed through analogy with plane expansion due to plane n-doping. This also holds for the C- phase, which in addition displays a similarly shortened c-axis as indicated in the laser pulsing experiments. A main piece of information concerning stability vs. transformations pertains to the analysis of E=280kJ for the transition O32/O24. This high activation energy lets a parent of this exotic V+, presumably subperoxide based material survive SQ in reasonably ordered state at temperatures as high as 950K. At this temperature y is fixed around ideal 6.44, so that it can be SQ and then be reordered around 670K where it doesn't readily absorb more oxygen.

We will now introduce arguments leading to the different stability of O32 and O243 versus transitions to O24. To begin with, the transition of O42 to O243 with rising temperature is considered to take place in infinitely adaptable if somewhat disordered superstructures, O42 being stable in its pure form around 260K and O32 being documented after shot quenching from 673K. We hold that disorder will play a minor role, given the energetic cost of strong O-O repulsion potentials. Accordingly chain length of (4) decreases systematically from infinity to 0 by simple shifts in the O position. As an example chain length is shortened by about half on moving one O out of a chain of (4) into the chain of (2) as the temperature is raised beyond the characteristic temperature for the given chain length. In the process the chain of (4) is now locally transformed into environment (4)O(3)(3)O(4) and the transferred O creates a new environment within the corresponding chain of (2) of (2)(3)O(3)(2). We hold the latter to be doping neutral so that the loss of (4) will systematically decrease Tc immediately after SQ. On aging at 298K Tc will recover within hours in the well-studied O24. Our SQ method with its t=0.1s has extended the range of observation up to 470K while effects saturated at higher temperatures due to the limits of this time window.

However, for SQ from Ta>600K new phenomenology with V+ effects arises, maximizing near 673K. It is clear that these effects survive the higher temperatures because of an inherently different electronic state, increasing E to 280K. We hold subperoxide formation responsible for this different nature. In particular, we consider a critical local environment of two adjacent units (3)O(3) in the form (3)O(3)(3)O(3) to be responsible for the first generation of $O^-$ on the apical location around (3). This starts to form at the critical Ta=600K on the ladder concept. At this point a new type of (3) is becoming associated with plane n-doping and it is in competition with plane p-doping (4). In this range Tc<4.2K. When the ladder structure has practically excluded (4) at 673K, the V+ material can attain the Tc=100K level.

Accordingly conventional Tc aging effects at room temperature are seen as final steps within a larger phenomenology based primarily on several high temperature steps of intrinsically slow metal ligand charge rearrangement under thermal stress.

**4. Conclusion**

YBa2Cu3Oy near O-half filling shows several high temperature modifications that are intrinsically slow to transform within shot quenching techniques. We conclude that this has to do with the need for electronic rearrangement and propose that the unusual electronic states are based on $O^-$ in the form of subperoxides. This is in line with a stability belt for peroxides around this temperature window. High temperature modifications are distinct by their unusual properties such as n-doping, multiples of Tc, expanded cell volumes or axial ratios, denoted as V+, C- or ETS. By contrast, on slow cooling to near room temperature, the conventional cell volume contracted V- type is based on O24 with plane self-doping of p-type by the highly oxidized metal chain Cu (4).

Transitions between these qualitatively different self-doped types, based on either $O^-$ or metal, are much slower (observable transitions in the 1000K range indicate E to 300kJ levels) than purely structural ones from O243 to O24 (E at 92kJ level observable with SQ up to 470K). We argue that the lowest belt of subperoxide stability starts at 600K and carries apical $^cO^-$ on (3) with critical condition of 2 adjacent (3)O(3), increasing axial ratio c/p. At 670K cell volume maximizes in O32 of V+P+ type with a Tc=100K level due to n-doped plane charge of c=0.22. Stability of O32 extends to 950K. At 970K, transformations can lead to amorphous material due to the competition with a higher temperature C- modification. It is argued that C- is based on intrachain charge compensation involving $^cO^-$ and metal. C- is seen as related to a yet to be identified phase with ETS, which is proposed to exhibit subperoxidic $^cO^-$ on both apical sites connected with (2), strongly contracting its c-axis.

Table 1. Transformations with different kinetic in $YBa_2Cu_3O_y$ near O-half filling between the following distinct phases with different temperatures of stability: T32 (1100K), TC- (1050K), amorphous (970K), O3 (950-670K), O342 (670-600K), O423 (600-260K).
Transformation kinetic gates further development on cooling or re-oxygenation (rox). V+p stands for parent high temperature modification, which can develop their intriguing electronic and structural properties through thermal stress on SQ.
A main difference pertains to T23 (majority (2)) and O3 parentage (majority (3) with minor amounts of (2) implied), gating into orthorhombics based on (4) and (3) respectively. T23, obtainable on annealing (a) at >1100K (y<6.25) can develop, on slow oxygenating cooling (SC), into conventional O24 of P-V- type (Tc=50K level). O24 is based on oxidative self-doped $(5)^{Ox}$ through charge equilibration with $(4)^{Red}$. By contrast, shot quenching (SQ) y>6.4 (673- 950K with ideal y=6.44) can develop into phenomenology of O3, containing doped oxidized (Ox) c-axis (apical) $O^-$. This $^cO^{Ox}$ on (3) can lead to P+V+ effects with charge equilibration on $(5)^{Red}$ (Tc=100K level).
A further distinction pertains to transformations of O24 and O3 derived materials. When O24 is held up to 600K it develops O243 with non-doping (3), which is transforming back to O24 with E=92kJ and $A=1.0x10^{-12}$s after SQ to 298K. By contrast varieties based on $O^-$ such as O3, obtained by annealing at 950K transform to O24 with up to 280kJ. Mixed O324, obtained by SQ between 600-670K, transform their non-doping (3) with E=92kJ and their doping $(3)O^{Ox}$ with E=280kJ. The latter transformation is slow on account of its electronic rearrangement by comparison with the mainly structural one within O243/O24.
C- phase is stable at 1050K for hours, indicating E>310kJ. At 970K amorphous materials obtain, indicating perhaps an inherent instability to decomposition including peroxides such as $BaCuOx$ or $BaO_2$ with higher overall O content (e.g. 8.5).
$O^-$ phenomenology, based on $(5)^{Red}$ with P+ and balancing apical $^cO^{Ox}$, can be seen as creating new superconductors with increased Tc, such as V+ and ETS. It can also result in Tc=0 as in the C- phase or transitional types between V- and V+. C- is explained on internal charge compensation on chains involving $O^-$. C- is seen as a precursor of ETS, where 2 apical $^cO^{Ox}$ are assumed as coordinating (2) and charge equilibrating with $(5)^{Red}$.

| Structure, Tc(K) | Ta (K) | final doped | V ($Å^3$) | E (kJ) |
|---|---|---|---|---|
| T23V0/O24V-, 50 | 1070 SC | $(5)^{Ox},(4)^{Red}$ | 173 | |
| C-/O34,0 | 1050 SC | $(5)^{2+},0.3(2)2^cO^-,0.7(4)^{2+}$ | 174.5 | >310 |
| ETS, 200* | | $(5)^{Red},(2)2^cO^{Ox}$ | | |
| Amorph | 970 SQ | | | |
| O3V+p/O24V- | 950 1h/SQ | | | 280 |
| O3V+p/O23V+ | 950 SQrox | $(5)^{Red},(3)O^{Ox}$ | 177 | 236 |
| O324V- p, 0 | 950 SQ | $(5)^{Ox},(3)O^{Ox},(4)^{Red}$ | | |
| O32V+, 100 | 673 SQ | $(5)^{Red}, (3)O^{Ox}$ | 177 | >140 |
| O423V-/O42V-, 50 | 473 SQ | $(5)^{Ox},(4)^{Red}$ | 173 | 92 |